\documentclass[useAMS,usenatbib]{mn2e}
\usepackage{graphicx,amssymb,amsmath,times}
\title{A possible correlation between the high energy electron spectrum and the cosmic-ray secondary-to-primary ratios}
\author[Satyendra Thoudam]{Satyendra Thoudam\thanks{E-mail: s.thoudam@astro.ru.nl} and J\"{o}rg R. H\"{o}randel\thanks{E-mail: j.horandel@astro.ru.nl}\\
Department of Astrophysics, IMAPP, Radboud University Nijmegen, P.O. Box 9010, 6500 GL Nijmegen, The Netherlands}

\begin{document}
\date{}
\pagerange{}
\maketitle
\label{firstpage}
\begin{abstract}
Recent observations of high energy cosmic-ray electrons by the Fermi-LAT and the HESS experiments between $20$ GeV and $5$ TeV have found that the energy spectrum closely follow a broken power-law with a break at around $1$ TeV. On the other hand, measurements of cosmic-ray secondary-to-primary ratios like the boron-to-carbon ratio seem to indicate a possible change in the slope at energies around $100$ GeV/n. In this paper, we discuss one possible explanation for the observed break in the electron spectrum and its possible correlation with the flattening in the secondary-to-primary ratios at higher energies. In our model, we assume that cosmic-rays  after acceleration by supernova remnant shock waves, escape downstream of the shock and remain confined within the remnant until the shock slows down. During this time, the high-energy electrons suffer from radiative energy losses and the cosmic-ray nuclei undergo nuclear fragmentations due to their interactions with the matter. Once the cosmic-rays are released from the supernova remnants, they follow diffusive propagation in the Galaxy where they further suffer from radiative or fragmentation losses.
\end{abstract}

\section{Introduction}
Few years ago, measurements of high energy cosmic-ray (CR) electrons above $10$ GeV were performed mainly by balloon-borne experiments (Kobayashi et al. 2004 and references therein). Their measurements showed that the electron spectrum follow a power-law behaviour of the form $E^{-\Gamma}$ with the index $\Gamma\approx 3.2$ without any significant features up to energies around $2$ TeV. However, recent measurements made by the ATIC balloon experiment had found a sharp peak at $E\approx 600$GeV (Chang et al. 2008). But, this feature was not later detected by two other experiments, the spaced-borne Fermi-LAT and the ground-based HESS experiments (Abdo et al. 2009, Aharonian et al. 2008b, 2009) which measured the energy spectrum in the range of $20\mathrm{GeV}-5\mathrm{TeV}$. Their combined spectrum can be closely represented by a broken power-law with spectral indices $\Gamma_1\approx 3$ and $\Gamma_2\approx 4$ for energies below and above $1$ TeV respectively. In standard CR propagation studies assuming a homogeneous source distribution, one way to explain such a spectral behavior is to assume that the electron source spectrum follow a power-law behavior  with an exponential cut-off of the form exp$(-E/E_c)$ with $E_c\approx 1$TeV. Radio and X-ray observations do support such a form of electron spectrum inside supernova remnants (SNRs), but with cut-off energies as high as $E_c\approx 80$ TeV as suggested by the study of Reynolds $\&$ Keohane 1999 considering magnetic field strengths of $10\mu$G inside the remnants. However, for young SNRs where magnetic field amplification seems to occur, the field strengths can reach values even more than $100\mu$G (V\"{o}lk et al. 2005) and the maximum energies for the electrons can be strongly limited by radiative losses to values less than $\sim 10$ TeV. 

The observed break can also be an effect of inhomogeneous distribution of CR sources contributing to the electrons at higher energies (Atoyan et al. 1995, Kobayashi et al. 2004, Blasi 2009). This is because high energy electrons cannot travel far distances in the Galaxy due to their faster radiative cooling. Therefore, it is possible that most of the TeV electrons that we measure in the Solar System are produced by few young nearby sources. Then, the general assumption of a continuous source distribution may break down and a more reasonable treatment would be to take into account a discrete source distribution. A common method to do this is to use the idea of separation of distant and nearby sources, i.e., assuming a continuous distribution for the distant sources and considering known SNRs or pulsars as the nearby discrete sources. One of the main uncertainties involved in such studies can be the effect of missing sources due to detection biases. Moreover, even for the known sources, the lack of precise informations about the source parameters like the distance, age and the spectral informations can lead to strong uncertainties in the high energy electron spectrum. Recently Delahaye et al. 2010 presented a detailed study using this method where they included all the known SNRs and pulsars located within $2$ kpc from the Earth and they found that the total high energy spectrum from the nearby sources strongly depends on several source parameters including the assumed cut-off energy. They also showed that the cut-off energy should be somewhere around a few TeV in order to explain the break in the observed data. An alternative explanation for the break, as will be shown in this paper, is that after acceleration the high energy electrons might have suffered significant  losses within the sources itself before they are released into the Galaxy.

On the other hand, measurements of CR secondary-to-primary (s/p) ratios like the boron-to-carbon (B/C) up to $E\approx 100$ GeV/n by several independent experiments (see the experiments listed in Stephens $\&$ Streitmatter 1998) show that above $1$ GeV/n, the ratio decreases with energy as $\sim E^{-\delta}$ with the index $\delta\approx 0.6$ implying an energy dependent CR propagation path length in the Galaxy. However, these measurements seem to indicate a possible hardening at higher energies (see also Swordy et al. 1990). Recent data from the CREAM balloon-borne experiment also show flattening at energies around $1$ TeV/n (Ahn et al. 2008). One way to explain this is to assume that CRs traverse some minimum amount of matter of approximately $0.3$ g cm$^{-2}$ in the Galaxy (Ave et al. 2009). This implies an energy dependent escape path length of CRs up to some certain energy and beyond that energy, the path length becomes constant in energy. This view is, in fact, supported by the observed CR anisotropy which remain almost constant above around $100$ GeV energies. But such model need to assume a break in the source spectrum in order to explain the observed CR spectrum which closely follow a pure power-law without any break up to the knee. Another possible explanation is that some amount of secondaries might be produced inside the sources due to the nuclear interaction of the primaries with the matter (Berezhko et al. 2003). Under such models, the CR anisotropy is expected to increase with energy as $E^\delta$ which does not agree well with the observed data. However, it is quite possible  that at higher energies the anisotropy might not be determined by the global diffusion leakage of CRs from the Galaxy. Rather, it might be the effects of high energy CRs coming from nearby sources because of their faster diffusion in the Galaxy (Ptuskin et al. 2005, Thoudam 2007). One more possible explanation which we will not discuss here is the possible reacceleration of some fraction of the background CRs by strong SNR shocks while propagating in the Galaxy (Wandel et al. 1987, Berezhko et al. 2003).

In this paper, we present one possible explanation for the observed break in the electron spectrum at around $1$ TeV and its possible correlation with the flattening in the B/C ratio at energies above around $100$ GeV/n. In our model, we assume that CRs are accelerated by SNR shock waves by diffusive shock acceleration (DSA) mechanism (Bell 1978, Blandford $\&$ Eichler 1987). Under DSA theory, suprathermal particles in the tail of the Maxwell-Boltzmann distribution present in the interstellar medium (ISM) are injected into the supernova shock front. They are then reflected back and forth several times across the shock by the magnetic turbulence generated on either side of the shock and in each crossing, the particles gain energy by first order Fermi acceleration. In a simple planar shocks model, such a mechanism naturally leads to a power-law spectrum of the form $E^{-\Gamma}$ with $\Gamma=2$ for strong shocks which is in good agreement with radio observations of several SNRs (Green 2009).

One important consideration of the standard DSA theory is that most of the particles do not escape upstream of the shock, they can only escape downstream mainly due to advection by the bulk flow and remain confined within the remnant due to the strong magnetic turbulence generated by the CRs themselves. Though it is still not fully understood, it is generally considered that the particles remain inside the remnant until the shock remains strong. During the confinement period, the high energy electrons may suffer from radiative energy losses while the nuclear components suffer from nuclear fragmentations. At later stages when the shock slows down and does not efficiently accelerate the CRs, the magnetic turbulence level goes down and the particles can no longer be confined effectively within the remnant. At such stage, all the particles present inside the remnant escape and they are injected into the ISM. For a typical ISM density of $1$ H cm$^{-3}$, this happens at around $10^5$ yr after the supernova explosion (Berezhko $\&$ V\"{o}lk 2000). The scenario described above might be true mostly for the lower energy particles because it is quite possible that some of the highest energy particles may escape the remnant already at the start of the sedov phase itself due to their faster diffusion. Such an energy dependent escape of particles has been discussed in a number of literatures (Ptuskin $\&$ Zirakashvili 2005, Gabici, Aharonian $\&$ Casanova 2009). Recent high energy $\gamma$-ray observations of SNRs associated with molecular clouds also suggest that some of the high energy particles might have already escaped the remnant at much early times (Aharonian et al. 2007, Aharonian et al. 2008a). But, as mentioned above, it is still not clear how and when exactly the particles escape the remnant. For simplicity, we consider an energy independent scenario for our present work and assume that all the particles are released into the ISM at some characteristic time after the supernova explosion.

Once the CRs are released into the ISM, we assume that they undergo diffusive propagation in the Galaxy where the electrons again suffer from radiative losses and the nuclei from nuclear spallation with the interstellar matter. Therefore, in our model the CR spectra that we finally observe in the Solar System are modified from their original source spectrum (the one generated by the SNR shocks) due to the various interactions or energy loss processes occurring not only during their propagation in the Galaxy, but also within the SNRs itself.

Our paper is planned as follows. In section 2, we present our calculations for the CR spectra inside the SNRs and in section 3, we calculate the spectra in the Galaxy. Then, in section 4, we compare our results with the observed data and also give a short discussions about our results. Finally in section 5, we give a brief conclusion of our study.

\section {CR spectra within the SNRs}
We assume that CR acceleration by SNR shock waves begins at the time of the supernova explosion itself and the CRs then escape downstream of the shock and remain confined within the remnant. We further assume that at some later stage of the SNR evolution characterize by age $t=T$ when the shock slows down, particle acceleration stops and also all the CRs present inside the remnant are released into the ISM. In our study, we do not consider the expansion of the remnant and hence neglect the effect of adiabatic energy losses and other related effects. Detailed study including the evolution of the remnant, the weakening of the shock with time and the possible energy dependent escape of CRs will be presented elsewhere.

\subsection {High energy electrons}
Under the assumption that the acceleration time of CRs is much less than their confinement time within the SNR, the CR electron spectrum inside an SNR can be described by the following equation,
\begin{equation}
\frac{\partial N_e}{\partial t}-\frac{\partial}{\partial E}\lbrace b(E) N_e\rbrace =q_e
\end{equation} 
where $N_e(E,t)$ is the total number of electrons of kinetic energy $E$ present within the SNR at time $t$, $b(E)=-dE/dt$ is the energy loss rate and $q_e(E)$ is the source term which, in our case, is the rate at which CR electrons are injected downstream of the shock. We believe that Eq. (1) represents a valid approximation at least for energies up to around $10$ TeV for which the typical acceleration timescale is less than $\sim 100$ yr (Sturner et al. 1997). In Eq. (1), we consider the energy loss of the electrons to be due to synchrotron and inverse compton interactions which is true for energies $E\gtrsim 10$ GeV. Therefore, we take in the Thompson regime
\begin{equation}
b(E)=aE^2
\end{equation}
with $a=1.01\times 10^{-16}(w_{ph}+w_B)$ GeV$^{-1}$ s$^{-1}$, where $w_{ph}$ and $w_B$ are the energy densities of the radiation fields and the magnetic field respectively in eV cm$^{-3}$. With this, we define the radiative energy loss timescale for the electrons as $t_{loss}(E)=1/(aE)$. Now, for the source spectrum taken as power-law of the form,
\begin{equation}
q_e(E)=k_eE^{-\Gamma}
\end{equation}
the solution of Eq. (1) at time $t$ can be written as (Kardashev 1962, Gratton 1972)
\begin{eqnarray}
N_e(E,t)=\frac{k_eE^{-(\Gamma+1)}}{(\Gamma-1)a}\left[1-\left(1-aEt\right)^{\Gamma-1}\right], \; \mathrm{for}\;E<E_0\nonumber \\
=\frac{k_eE^{-(\Gamma+1)}}{(\Gamma-1)a}, \qquad\qquad\qquad\qquad\quad\;\mathrm{for}\; E\geq E_0
\end{eqnarray}
where $E_0=1/(at)$ is the energy at which the energy loss time $t_{loss}$ becomes equal to time $t$. Eq. (4) shows that for energies $E\ll E_0$, the electron spectrum still reflects the source spectrum because of their large $t_{loss}$, i.e. $N_e(E)\propto E^{-\Gamma}$. However, for electrons with energies $E\gg E_0$, their spectrum is quite steeper due to their faster energy loss rate and follows $N_e(E)\propto E^{-(\Gamma+1)}$. 

\subsection {CR nuclei}
 For the CR primary nuclei, the spectrum inside the SNR can be described by
\begin{equation}
\frac{\partial N_p}{\partial t}+\eta^{\prime}c\sigma_p N_p =q_p
\end{equation}
where $N_p(E,t)$ is the total number of particles of kinetic energy per nucleon $E$ at time $t$, $\eta^{\prime}$ is the matter density in the SNR, $c$ is the velocity of light and $\sigma_p$ is the spallation cross-section of the primary nuclei assumed to be independent of energy. In Eq. (5), we assume that the source term $q_p(E)=k_pE^{-\Gamma}$ has the same index as those of the electrons.

The solution of Eq. (5) can be obtained as
\begin{equation}
N_p(E,t)=\frac{q_p(E)}{\eta^{\prime}c\sigma_p}\left[1-e^{-\eta^{\prime}c\sigma_p t}\right]
\end{equation}
For time much less than the nuclear spallation time, i.e. $t\ll 1/(\eta^{\prime}c\sigma_p)$, Eq. (6) becomes $N_p(E,t)\approx q_p(E) t$.

During the time when primary CRs are confined within the SNRs, they interact with matter and produce secondary nuclei of almost the same kinetic energy per nucleon as their primaries. These secondaries can also be described by an equation similar to Eq. (5) by replacing the source term by $\eta^{\prime}c\sigma_{ps}N_p(E,t)$ where $\sigma_{ps}$ is the total fragmentation crossection of primary to secondary.  The solution for these secondaries is then obtained by
\begin{equation}
N_s(E,t)=\frac{\sigma_{ps}}{\sigma_p}\frac{q_p(E)}{\eta^{\prime}c\sigma_s}\left[1-e^{-\eta^{\prime}c\sigma_s t}+ F\right]
\end{equation}
where,
\begin{equation*}
F=\frac{\sigma_s}{(\sigma_s-\sigma_p)}\left(e^{-\eta^{\prime}c\sigma_s t}-e^{-\eta^{\prime}c\sigma_p t}\right)
\end{equation*}
In Eq. (7), $\sigma_s$ represents the spallation cross-section of the secondary nuclei. Now, taking $t=T$ which is the CR confinement time inside the SNR, we can use Eqs. (4), (6) $\&$ (7) to calculate the spectrum of CR electrons and the spectra of primary and secondary CR nuclei finally injected into the ISM from a single SNR. Then, knowing the rate of supernova explosion per unit volume in the Galaxy, we can calculate the rate at which CRs are injected per unit volume in the Galaxy. 

\section {CR spectra in the Galaxy}
CRs after escaping from the SNRs undergo diffusive propagation in the Galaxy due to scattering either by magnetic field irregularities or by self excited Alfven and hydromagnetic waves. For the present work, we assume the diffusion region as a cylindrical disk of infinite radius with finite half-thickness $H$ and that the sources as well as the matter are distributed uniformly and continuously in the Galactic disk with half-thickness $h$ and radius $R$, where $R\gg H\gg h$. The details of the geometry are described in Thoudam 2008. During the propagation, high energy electrons interact with the background radiation and the magnetic fields and lose their energies. On the other hand, CR nuclei undergo nuclear spallation interactions with the interstellar matter and produce lighter nuclear species.

As already mentioned in section 1, the assumption of a continuous source distribution may not be fully appropriate for high energy electrons particularly those in the TeV region because of their faster energy loss rate. Electrons with energies greater than $1$ TeV cannot travel distances more than $\sim 1$ kpc in the Galaxy through diffusive propagation before they lost all their energies. Therefore, high energy electrons from distant and old sources may not reach the Earth effectively and TeV electrons that we observe can be mostly dominated by those produced by few young local sources. The effect of this can be that the spectrum at high energies can be quite complex because of its strong dependence on the local source parameters (Delahaye et al. 2010). Moreover, it may even show up features related to the stochastic nature of the local sources in space and time (see e.g., Pohl $\&$ Esposito 1998).

In our study, our main focus is to explore the possibility of explaining the break in the electron spectrum as an effect of CR confinement in the downstream region of SNRs and at the same time, following the standard model of CR propagation in the Galaxy. Such an effect of confinement within the sources are generally not considered in CR propagation studies and we believe that they may exist if SNRs are the main sources of galactic CRs. Therefore, we do not intend to focus on a detailed source distribution and in what follows, we adopt the continuous and stationary source distribution for calculating the CR spectrum in the Galaxy.

\subsection {High energy electrons}
Under the diffusion model, the propagation of high energy electrons in the Galaxy can be described by
\begin{equation}
\nabla\cdot(D\nabla n_e)+\frac{\partial}{\partial E}\lbrace b(E)n_e\rbrace=-Q_e
\end{equation}
where $n_e(\textbf{r},E)$ is the electron number density in the Galaxy, $D(E)$ is the diffusion coefficient which is assumed to be constant through out the Galaxy and $Q_e(\textbf{r},E)$ is the source term given by $Q_e(\textbf{r},E)=\Re N_e(E,T)\delta (z)$. Here, $N_e(E,T)$ is given by Eq. (4) which represents the total amount of CR electrons liberated by an SNR and $\Re$ denotes the rate of supernova explosion per unit surface area on the Galactic disk. As already mentioned, high energy electrons cannot travel large distances because of their large energy loss rate. Therefore, their equilibrium spectrum is not much affected by the presence of the halo boundary $H$ and we solve Eq. (8) without imposing the finite boundary conditions for our study. The solution at the position of the Earth ($z=0$) is then obtained as
\begin{equation}
n_e(E)=\frac{\Re}{2\sqrt{\pi}b(E)}\int_E^\infty dE^{\prime} N_e(E^{\prime},T)\left[\int_E^{E^{\prime}}\frac{D(u)}{b(u)}du\right]^{-1/2}
\end{equation}
For the diffusion coefficient $D(E)\propto E^\delta$, Eq. (9) shows that
\begin{equation}
n_e(E)\propto\frac{1}{b(E)}\int_E^\infty dE^{\prime} \frac{N_e(E^{\prime},T)}{\sqrt{E^{\delta-1}-E^{\prime \delta-1}}}
\end{equation}
From Eq. (10), we see that for a power-law source spectrum such as $N_e(E)\propto E^{-\gamma_e}$, we get $n_e(E)\propto E^{-\gamma_e-1+\beta}$ where $\beta=(1-\delta)/2$. Thus, the equilibrium electron spectrum in the Galaxy is modified mainly by the radiative energy losses with a small correction $\beta$ arising due to the diffusive propagation effect.

\subsection {CR nuclei}
The steady state transport equation for primary CR nuclear species in the Galaxy, neglecting the effect of energy losses due to ionization, is given by
\begin{equation}
\nabla\cdot(D\nabla n_p)-2h\eta c\sigma_p\delta(z)n_p=-Q_p
\end{equation}
where $n_p(\textbf{r},E)$ is the density of primary CRs of energy per nucleon $E$, $\eta$ is the matter density in the ISM and $Q_p(\textbf{r},E)=\Re N_p(E,T)\delta (z)$ is the source term where $N_p(E,T)$ is given by Eq. (6). Eq. (11) is solved by imposing proper boundary conditions as described in detail in Thoudam 2008 and the solution at $z=0$ is given by  
\begin{equation}
n_p(E)=\frac{R\Re N_p(E,T)}{2D}\int^\infty_0\frac{J_1(KR)dK}{\left[K\mathrm{coth}(KH)+\frac{2h\eta c\sigma_p}{2D}\right]}
\end{equation}
where $\mathrm{J_1}$ is the Bessel function of order 1.

Primary nuclei during their propagation in the Galaxy interact with matter in the Galaxy and produce secondary nuclei. The equilibrium spectrum for these secondaries $n_s(\textbf{r},E)$ in the Galaxy can be described by
\begin{equation}
\nabla\cdot(D\nabla n_s)-2h\eta c\sigma_s\delta(z)n_s=-Q_s
\end{equation}
where the source term $Q_s(\textbf{r},E)=2h\eta c\sigma_{ps}n_p(\textbf{r},E)\delta(z)$. There is also an additional component of the secondaries in the Galaxy due to their production inside the SNRs as discussed in section 2.2. The equilibrium spectrum $n_s^\prime$ for this additional component can also be described by an equation similar to Eq. (13) with the source replacing by $Q^\prime_s(\textbf{r},E)=\Re N_s(E,T)\delta (z)$, where $N_s(E,T)$ is given by Eq. (7). The spectra of these two secondary components at the position of the Earth can be obtained respectively as
\begin{equation}
n_s(E)=2h\eta c\sigma_{ps}n_p(E)\frac{R}{2D}\int^\infty_0\frac{J_1(KR) dK}{\left[K\mathrm{coth}(KH)+\frac{2h\eta c\sigma_s}{2D}\right]}
\end{equation}
and
\begin{equation}
n_s^\prime(E)=\frac{R\Re N_s(E,T)}{2D}\int^\infty_0\frac{J_1(KR)dK}{\left[K\mathrm{coth}(KH)+\frac{2h\eta c\sigma_s}{2D}\right]}
\end{equation}

Having calculated the spectra for the primaries and the secondaries in the Galaxy, the overall s/p ratio can be obtained by simply taking the ratio $(n_s+n_s^{\prime})/n_p$. Note that in the standard model of CR propagation in the Galaxy where secondary nuclei are assumed to be produced only in the ISM, the ratio is given by $n_s/n_p$.

\section{Results and discussions}
For our calculations, we choose $R=16$ kpc, $H=5$ kpc, $h=200$ pc and the rate of supernova explosion as $\Re=25$ Myr$^{-1}$ kpc$^{-2}$ (Grenier 2000). This corresponds to a total supernova explosion rate of $\sim 1/50$ yr in our Galaxy. The total energy density of the radiation fields is taken as $w_{ph}=w_{MBR}+w_{op}+w_{FIR}$ where $w_{MBR}$,  $w_{op}$ and $w_{FIR}$ are the energy densities of the microwave background, the NIR-optical  and the FIR radiation fields respectively. We take $w_{MBR}=0.25$ eV cm$^{-3}$, $w_{op}=0.5$ eV cm$^{-3}$ (Mathis, Mezger $\&$ Panagia 1983), $w_{FIR}=0.2$ eV cm$^{-3}$ (Chi $\&$ Wolfendale 1991) and the magnetic field value as $B=6\mu$G (Beck 2001). We assume these values to be same for both in the Galaxy and in the SNRs. We take the ISM density as $\eta=1$ cm$^{-3}$ (Gordon $\&$ Burton 1976, Bronfman 1988) typical for our Galaxy and the CR diffusion coefficient in the Galaxy as (Thoudam 2008)
\begin{eqnarray}
D(E)=D_0\left(\frac{\xi}{E}\right)^{\delta},\quad\quad \mathrm{for}\;E<\xi \nonumber\\
\qquad =D_0\left(\frac{E}{\xi}\right)^{\delta},\quad\quad \mathrm{for}\;E>\xi
\end{eqnarray}
where $D_0=2.9\times 10^{28}$ cm$^2$s$^{-1}$ and $\delta=0.6$. For CR nuclear species, $E$ represents energy per nucleon and the particle rigidity $\rho$ (corresponding to $\xi$) for charge $Z$ and mass number $A$ is taken as $\rho=A\xi/Z=3$ GV. For electrons, $E$ in Eq. (16) represents the kinetic energy and $\xi$ is directly taken as $3$ GeV.

Finally, we take the source spectral index $\Gamma=2.2$ and the CR confinement time within the SNRs as $T=1.2\times 10^5$ yrs. Then, using the solutions for the CR spectra derived in section 3, we calculate the CR electron spectrum and the B/C ratio in the Galaxy and compare them with the observed data. In Fig. (1), we plot our calculated electron spectrum normalized to the data reported by the Fermi and the HESS experiments at $50$ GeV. This corresponds to approximately $0.3\%$ of the total kinetic energy of the supernova explosion, taken as $E_{SN}=10^{51}$ ergs, converting into the CR electrons above $1$ GeV. In Fig. (2), we plot the B/C ratio in the Galaxy where the solid line represents the ratio expected under the standard model which assumes  boron production entirely only in the ISM, the dotted and the dot-dashed lines represent the results of our model which also include additional boron production inside the SNRs for $\eta^\prime=1$  cm$^{-3}$ and $2$ cm$^{-3}$ respectively. In our calculation, we consider that boron nuclei are produced due to the spallation of carbon and oxygen nuclei and we take the oxygen to carbon (O/C) source abundance ratio as 1.4. The data in Fig. (2) includes the compilation given in Stephens $\&$ Streitmatter 1998 and also the recent data from the CREAM experiment (Ahn et al. 2008). From the figures, we see that our model explains quite well the observed break in the electron spectrum at $E\sim 1$ TeV without invoking any exponential cut-off in the source spectrum and also it explains the observed flattening in the B/C ratio above $\sim100$ GeV/n. 

\begin{figure}
\centering
\includegraphics*[width=0.33\textwidth,angle=270,clip]{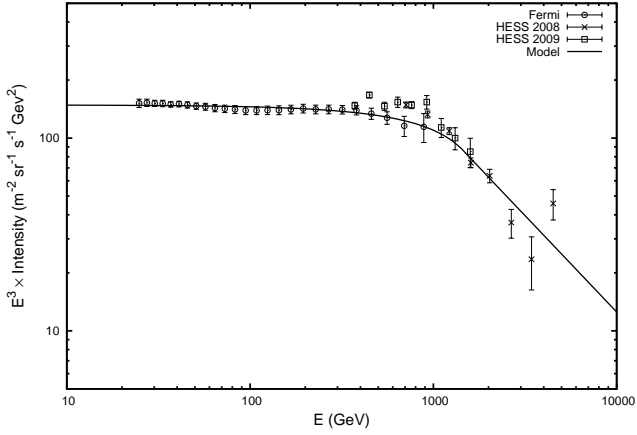}
\caption{\label {fig1} CR electron spectrum at the position of the Earth. The solid line is the result of our calculation using Eq. (9) normalized to the data at $50$ GeV. The data are taken from the Fermi (Abdo et al. 2009) and the HESS (Aharonian et al. 2008b, 2009) experiments. This normalisation corresponds to $\sim 3\times 10^{48}$ ergs of the supernova explosion energy converting into CR electrons above $1$ GeV. Our model parameters: $T=1.2\times 10^5$ yrs, $\Gamma=2.2$, $\delta=0.6$, $\Re=25$ Myr$^{-1}$ kpc$^{-2}$ and $E_{SN}=10^{51}$ ergs. Other model parameters: $w_{MBR}=0.25$ eV cm$^{-3}$, $w_{ph}=0.5$ eV cm$^{-3}$, $w_{FIR}=0.2$ eV cm$^{-3}$ and $B=6\mu$G, which are taken to be same both in the Galaxy and in the SNRs.}
\end{figure}

\begin{figure}
\centering
\includegraphics*[width=0.33\textwidth,angle=270,clip]{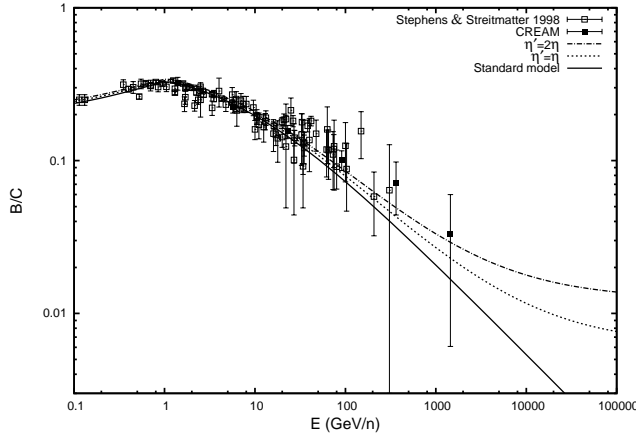}
\caption{\label {fig2} B/C ratio at the position of the Earth. The solid line is the ratio under the standard model where boron is assumed to be produced only in the ISM. The results of the present model, where boron production inside the sources are included, are shown for two values of matter density inside the remnants $\eta^\prime=1$ cm$^{-3}$ (dotted line) and $2$ cm$^{-3}$ (dot-dashed line) respectively. The ISM density is taken as $\eta=1$ cm$^{-3}$. The model parameters used are the same as in Fig. (1).}
\end{figure}
Our results can be understood as follows. We see from section 3.1 that the spectrum of high energy electrons in the Galaxy follow $n_e(E)\propto E^{-\gamma_e-1+\beta}$, where $\gamma_e$ is the index of electron injection spectrum into the Galaxy and $\beta=(1-\delta)/2$ is a small correction due to propagation. For the diffusion index $\delta=0.6$, we get $\beta=0.2$. In our model, $\gamma_e$ is the spectrum of CR electrons produced by the SNRs (see section 2.1) and is given by $\gamma_e=\Gamma$ for energies $E\ll E_0$ and $\gamma_e=\Gamma+1$ for $E\gg E_0$. Therefore, for $\Gamma=2.2$ we get,
\begin{eqnarray}
\gamma_e=2.2,\qquad \mathrm{for}\; E\ll E_0\nonumber\\
=3.2,\qquad \mathrm{for}\; E\gg E_0
\end{eqnarray}
which give the corresponding high energy electron spectrum in the Galaxy as $n_e(E)\propto E^{-3}$ and $E^{-4}$ respectively. The energy at which the spectral break occurs depend both on the CR confinement time within the remnants and also on the total energy density $w_{ph}$ of the radiation and the magnetic fields present in the SNRs as $E_0=1/(aT)$. For the values of $T$, $w_{ph}$ and $w_B$ adopted in the present work, we obtain $E_0=1.4$ TeV.

The spectra of high energy primary nuclei in the Galaxy follows $n_p(E)\propto E^{-\gamma_p-\delta}$, where $\gamma_p$ is the primary injection spectrum in the Galaxy. For the secondary nuclei, those which are produced in the ISM follow a spectrum given by $n_s(E)\propto E^{-\gamma_p-2\delta}$ in the Galaxy whereas, those produced due to the interaction of the primaries within the remnants follow $n_s^\prime(E)\propto E^{-\gamma_s-\delta}$, where $\gamma_s$ is the secondary injection spectrum in the Galaxy. We see from section 2.2 that both the primary and the secondary injection spectra in the ISM have the same index as the intrinsic source spectrum, i.e. $\gamma_p=\gamma_s=\Gamma$. Thus, $n_s$ is steeper than $n_s^\prime$ by $E^{-\delta}$. Therefore, the overall s/p ratio follows,
\begin{eqnarray}
\frac{(n_s+n_s^\prime)}{n_p}\propto \frac{n_s}{n_p}\propto E^{-\delta},\qquad\qquad \mathrm{for}\; E\ll E_a\nonumber\\
\qquad \propto \frac{n_s^\prime}{n_p}= \mathrm{constant},\qquad \mathrm{for}\; E\gg E_a
\end{eqnarray} 
where $E_a$ is the energy at which the amount of secondaries produced in the ISM equals those produced within the SNRs, i.e. at $n_s=n_s^\prime$. For our results shown in Fig. (2), we get $E_a=7.5$ TeV/n and $2.5$ TeV/n for $\eta^\prime =1$ cm$^{-3}$ and $2$ cm$^{-3}$ respectively.
\begin{figure}
\centering
\includegraphics*[width=0.33\textwidth,angle=270,clip]{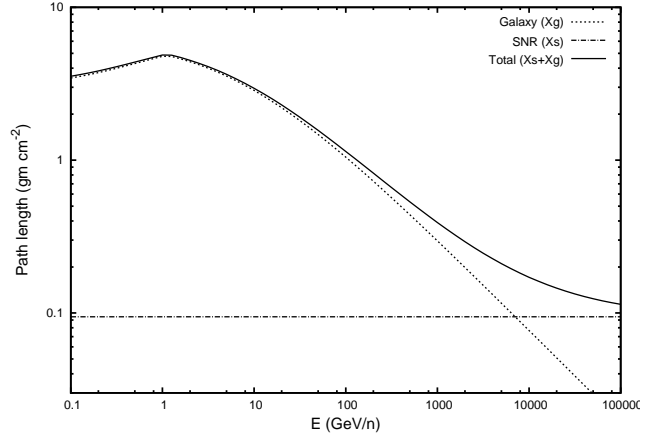}
\caption{\label {fig4} CR path length corresponding to the B/C ratio shown in Fig. 2. The dotted line represents the path length in the Galaxy $X_g$ and the dot-dashed line represents the path length within the SNRs $X_s$. The solid line represents the total path length ($X_s+X_g$) traversed during their total life time in the Galaxy. Model parameters: $T=1.2\times 10^5$, $\eta=1$ cm$^{-3}$, $\eta^\prime =1$ cm$^{-3}$ and $\delta=0.6$.}
\end{figure}

To better understand the relative production of CR secondaries inside the SNRs and in the Galaxy, it is good to compare the amount of matter traversed by the CRs during their confinement within the remnants to the amount traversed during their propagation in the Galaxy. For this, we calculate the path length $X$ as a function of s/p as,
\begin{equation}
X=\frac{m}{(\sigma_p-\sigma_s)}\mathrm{ln}\left[1-R_{sp}\frac{(\sigma_s-\sigma_p)}{\sigma_{ps}}\right]
\end{equation}
where $m$ denote the mass of hydrogen atom and $R_{sp}$ represent the s/p ratio. Using Eq. (19), we can calculate the CR path length for a given $R_{sp}$. For the path length calculation within the SNRs, $R_{sp}$ is given by the ratio $N_s/N_p$ taken at time $t=T$ (see section 2.2) and for the path length in the Galaxy, $R_{sp}=n_s/n_p$ from section 3.2. Their comparison corresponding to the B/C ratio shown in Fig. (2) is plotted in Fig. (3) where the dot-dashed line represents the path-length inside the SNRs $X_s$ and the dotted line that in the Galaxy $X_g$. Also shown in the figure by the solid line is the total path length ($X_s+X_g$) traversed by the CRs during their whole lifetime in the Galaxy. The average path length within the SNRs is found to be $X_s\approx 0.09$ gm cm$^{-2}$ for $\eta^\prime=1$ cm$^{-3}$ independent of energy whereas the path length in the Galaxy has a maximum value of $X_g\approx 5$ gm cm$^{-2}$ at around $1$ GeV/n which then decreases with energy as $X_g\propto E^{-\delta}$ because of the energy dependent escape of CRs from the Galaxy. From Fig. 3, we can also see that CRs with energies greater than around $7$ TeV/n traversed most of the matter within the source itself and its effect is seen in the B/C ratio in Fig. (2) already at energies around $1$ TeV/n. 

\section{Conclusions}
We present a simple model based on the considerations of DSA theory which can explain both the observed electron spectrum and the s/p ratios for a conservative set of model parameters. In our model, we assume that CRs after acceleration by SNR shock waves, escape downstream of the shock and remain confined within the remnant for some time before they are released into the ISM. During this time, CR electrons suffer from radiative energy losses while the nuclear species undergo nuclear fragmentations. For a magnetic field strength of $6\mu$G inside the SNRs and assuming a uniform and continuous distribution of SNRs in our Galaxy, we find that a CR confinement time of $1.2\times 10^5$ yr can produce the observed break in the electron spectrum at $\sim 1$ TeV. Moreover, the hardening in the available B/C data above $\sim 100$ GeV/n can also be explained if we assume the averaged matter density inside the SNRs to be $\sim 2$ cm $^{-3}$. Our results on the B/C ratio which are based on a simple model are very similar to those obtained  in Berezhko et al. 2003 for the case of secondary production inside SNRs calculated for the normal ISM density of $1$ cm$^{-3}$ and the CR confinement time of $10^5$ yr. Their results were calculated using  a detailed self consistent model of CR production inside SNRs.

Our model, in its present form, looks similar to the the nested leaky box model proposed by Cowsik $\&$ Wilson 1973, 1975. However, there are major differences in the basic assumptions between the two models. They assumed that CRs after acceleration spend some time in a cocoon-like region surrounding the sources where the primaries interact with the matter and produce secondaries. The residence time of CRs inside the cocoon was assumed to be energy \textit{dependent} and after they are released from the source region, they undergo an energy \textit{independent} propagation in the Galaxy. In their model, secondary production inside the cocoon dominates at lower energies and at higher energies, it is dominated by the production in the ISM. On the other hand, the basic idea of our model comes from our understanding of DSA theory inside SNRs. In our model, the CR confinement region as well as the confinement time are strongly related to the acceleration mechanism itself. Moreover, secondary production inside the remnant dominates only at higher energies while at lower energies, they are dominated mostly by those produced in the Galaxy. A more proper treatment of our model would be to perform a self consistent calculation of primary CR acceleration and their confinement, both of which are strongly related to the efficiency of CR scattering around the shocks by magnetic turbulence. Also, we should include the secondaries produced during the acceleration process along with those produced during the confinement period and if there, also their acceleration by the same shock waves which accelerate their primaries. Such a scenario of secondary acceleration has been considered to explain the rise in the positron fraction reported by the PAMELA experiment (Blasi 2009). Though it is beyond the scope of this paper to discuss this issue, it is worthwhile to mention in relation to our present model, that the relative contribution of the accelerated and the non-accelerated secondaries to their total spectrum strongly depends on the relative time their primaries spend in the acceleration region and in the downstream region (see e.g. Kachelriess et al. 2010).

We emphasize that under our present model, both the high energy electron spectrum and the s/p ratio depend strongly on the  CR confinement time inside the SNRs. Therefore, their data can be used to put constraints on the average confinement time provided that detailed informations about the magnetic field strengths and the matter densities inside the remnants are taken into account. Note that the value of the magnetic field strength of $6 \mu$G and also that of the matter density $\eta^\prime=(1-2)$ cm$^{-3}$ adopted in our present work can be different from those expected in SNRs. For SNRs expanding in an environment consisting of monoatomic ideal gas, their values are expected to be equal to the ISM values scaled by some constant factor which in the case of strong shocks is approximately equal to $4$. Moreover, these values can also be different for different SNRs and their true values can only be inferred from observations of radio or X-ray synchrotron emissions and thermal X-rays from SNRs (Bamba et al. 2003, Uchiyama et al. 2007). In future, we will include such details in our calculation and also try to extend our work to other secondary species like the CR anti-protons and the positrons. In addition, we will also include some important aspects which we have neglected in our present study like the evolution of the remnant, the weakening of the shocks and the energy dependent escape of CRs from the SNRs. Adding such aspects, particularly the energy dependent escape, can strongly affect our results. For instance, assuming an escape model which follows the same energy dependence as in the Galaxy can lead to disappearance of the flattening in the s/p ratio at higher energies.

\section*{Acknowledgments}
The authors would like to thank the anonymous referee for his/her constructive comments. ST would like to thank M. Pohl for useful discussions during NAC 2010 in Nijmegen. ST would also like to thank D. M\"{u}ller for carefully reading the first draft of this manuscript and for giving valuable suggestions.\\
\\
\\
\textbf{REFERENCES}\\
\\
Abdo, A. A., et al. 2009, Phys. Rev. Lett., 102, 181101\\
Aharonian, F. A., et al. 2007, A$\&$A, 467, 1075\\
Aharonian, F. A., et al. 2008a, A$\&$A, 481, 401\\
Aharonian, F. A., et al. 2008b, Phys. Rev. Lett., 101, 261104\\
Aharonian, F. A., et al. 2009, A$\&$A, 508, 561\\
Ahn, H. S., et al. 2008, Astropart. Phys., 30, 133\\
Atoyan, A. M., Aharonian, F. A., $\&$ V\"{o}lk, H. J., 1995, Phys. Rev. D, 52, 3265\\
Ave, M., Boyle, P. J., H\"{o}ppner, C., Marshall, J., $\&$ M\"{u}ller, D., 2009, ApJ, 697, 106\\
Bamba A., Yamazaki R., Ueno M., Koyama K., 2003, ApJ, 589, 827\\
Beck, R., 2001, Space Science Reviews, 99, 243\\
Bell, A. R., 1978, MNRAS 182, 147\\
Berezhko, E. G., Ksenofontov, L. T., Ptuskin, V. S., Zirakashvili, V. N., $\&$ V\"{o}lk, H. J, 2003, A$\&$A, 410, 189\\
Berezhko, E. G., $\&$ V\"{o}lk H. J., 2000, ApJ, 540, 923\\
Blandford, R., $\&$ Eichler, D., 1987, Physics Reports, 154, 1\\
Blasi, P., 2009, PRL, 103, 051104\\
Bronfman, L. et al. 1988, ApJ, 324, 248\\
Chang, J., et al. 2008, Nature, 456, 362\\
Chi, X., $\&$ Wolfendale, A. W., 1991, J. Phys. G, 17, 987\\
Cowsik, R., $\&$ Wilson, L. W., 1973, 13th ICRC, Denver, 1, 500\\
Cowsik, R., $\&$ Wilson, L. W., 1975, 14th ICRC, Munich, 1, 74\\
Delahaye, T., Lavalle, J., Lineros, R., Donato, F., Fornengo, N., 2010, A$\&$A, 524, 51\\
Gabici, S., Aharonian, F. A., $\&$ Casanova, S., 2009, MNRAS, 396, 1629\\
Gratton, L. 1972, ApSS, 16, 81\\
Green, D. A., 2009, BASI, 37, 45\\
Grenier I. A., 2000, A$\&$A, 364, L93\\
Gordon, M. A., Burton, W. B., 1976, ApJ, 208, 346\\
Kachelriess, M., Ostapchenko, S. $\&$ Tomas, R., 2010, arxiv: 1004.1118\\
Kardashev, N. S., 1962, Soviet Astronomy, 6, 317\\
Kobayashi, T., Komori Y., Yoshida K., $\&$ Nishimura J., 2004, ApJ, 601, 340\\
Mathis, J. S., Mezger, P. G., Panagia, N., 1983, A$\&$A, 128, 212.\\
Ptuskin, V. S., Jones F. C., Seo E. S., Sina R., 2005, 29th ICRC, Pune, OG.1.3\\
Ptuskin, V. S., $\&$ Zirakashvili, V. N., 2005, A$\&$A, 429, 755\\
Pohl, M., $\&$ Esposito, J. A., 1998, ApJ, 507, 327\\
Reynolds, S. P., $\&$ Keohane, J. W. 1999, ApJ, 525, 368\\
Stephens, S. A., $\&$ Streitmatter, R. E. 1998, ApJ, 505, 266\\
Sturner, S. J., Skibo, J. G., Dermer, C. D., Mattox, J. R. 1997, ApJ, 490, 619\\
Swordy, S. P., M\"{u}ller, D., Meyer, P., L'Heureux, J., $\&$ Grunsfeld, J. M. 1990, ApJ, 349, 625\\
Thoudam, S. 2007, MNRAS Letters, 380, L1\\
Thoudam, S. 2008, MNRAS, 388, 335\\
Uchiyama, Y., Aharonian, F. A., Tanaka, T., Takahashi, T., Maeda, Y., 2007, Nature, 449, 576\\
V\"{o}lk, H. J., Berezhko, E. G., Ksenofontov, L. T., 2005, A$\&$A, 433, 229\\
Wandel, A., Eichler, D., Letaw, J. R., Silberberg, R., $\&$ Tsao, C. H.  1987, ApJ, 316, 676\\
\end{document}